\documentclass[runningheads,a4paper]{llncs}
\usepackage[utf8]{inputenc}
\usepackage[T1]{fontenc}
\usepackage{graphicx}
\usepackage{wrapfig}
\usepackage{float}
\usepackage{amsmath}
\usepackage{amssymb}
\usepackage{hyperref}
\usepackage{listings}
\usepackage{xcolor}
\usepackage{url}
\usepackage{todonotes}
\author{Tsutomu Kobayashi\inst{1}\orcidID{0000-0002-8795-3183} \and Martin Bondu\inst{2} \and Fuyuki Ishikawa\inst{3}\orcidID{0000-0001-7725-2618}}
\urldef{\jaxaemail}\path|kobayashi.tsutomu@jaxa.jp|
\urldef{\suemail}\path|martin.bondu@etu.sorbonne-universite.fr|
\urldef{\niiemail}\path|f-ishikawa@nii.ac.jp|
\institute{
Japan Aerospace Exploration Agency, Tsukuba, Japan \\ \jaxaemail \and
Sorbonne University, Paris, France \\ \suemail \and
National Institute of Informatics, Tokyo, Japan \\ \niiemail
}
\title{Formal Modelling of Safety Architecture for Responsibility-Aware Autonomous Vehicle \\ via Event-B Refinement\thanks{The first author is supported by JSPS KAKENHI grant number 19K20249 and JST ERATO-MMSD (JPMJER1603) project. The third author is supported by JST MIRAI-eAI (JPMJMI20B8) project.}}

\newcommand{\precond}{\phi}
\newcommand{\propresp}{\alpha}
\newcommand{\scenario}{\mathcal{S}}
\newcommand{\env}{\mathsf{Env}}
\newcommand{\goal}{\mathsf{Goal}}
\newcommand{\safety}{\mathsf{Safety}}

\newcommand{\xsv}{x_{SV}}
\newcommand{\vsv}{v_{SV}}
\newcommand{\asv}{a_{SV}}
\newcommand{\xpov}{x}
\newcommand{\vpov}{v}
\newcommand{\apov}{a}
\newcommand{\lanes}{L}
\newcommand{\ctrl}{ctrl}
\newcommand{\vsvbcinit}{v_{BC0}}
\newcommand{\tcruisebc}{t_{BCCruise}}
\newcommand{\tbrakebc}{t_{BCBrake}}
\newcommand{\xLead}{x_{2}}
\newcommand{\tLCe}{t_{LCe}}

\newcommand{\xsvp}{p_x}
\newcommand{\vsvp}{p_v}

\newcommand{\xsvinit}{x_{SV0}}
\newcommand{\vsvinit}{v_{SV0}}
\newcommand{\xtgt}{x_{Tgt}}
\newcommand{\vmin}{v_{min}}
\newcommand{\vmax}{v_{max}}
\newcommand{\amax}{a_{max}}
\newcommand{\bmin}{b_{min}}
\newcommand{\bmax}{b_{max}}
\newcommand{\ac}{AC}
\newcommand{\bc}{BC}
\newcommand{\vLead}{v_{2}}
\newcommand{\tLC}{t_{LC}}

\newcommand{\gainv}{\phi}
\newcommand{\timetocruise}{timeToCruise}
\newcommand{\timetobrake}{timeToBrake}
\newcommand{\drss}{\mathsf{dRSS}}

\definecolor{labelcolor}{rgb}{0,0.4,0.8} % Labels, cyan
\definecolor{minuscolor}{rgb}{1,0.75,0.75}
\definecolor{pluscolor}{rgb}{0.75,1,0.75}

\newcommand{\blabel}[1]{\textcolor{labelcolor}{\small{\texttt{#1}}}}
\newcommand{\surroundrect}[1]{\framebox[1.1\width]{#1} }%\par}

\makeatletter
\let\old@lstKV@SwitchCases\lstKV@SwitchCases
\def\lstKV@SwitchCases#1#2#3{}
\makeatother
\usepackage{lstlinebgrd}
\makeatletter
\let\lstKV@SwitchCases\old@lstKV@SwitchCases

\lst@Key{numbers}{none}{%
    \def\lst@PlaceNumber{\lst@linebgrd}%
    \lstKV@SwitchCases{#1}%
    {none:\\%
     left:\def\lst@PlaceNumber{\llap{\normalfont
                \lst@numberstyle{\thelstnumber}\kern\lst@numbersep}\lst@linebgrd}\\%
     right:\def\lst@PlaceNumber{\rlap{\normalfont
                \kern\linewidth \kern\lst@numbersep
                \lst@numberstyle{\thelstnumber}}\lst@linebgrd}%
    }{\PackageError{Listings}{Numbers #1 unknown}\@ehc}}
\makeatother
\lstdefinelanguage{eventb}{
   morekeywords={Event,variables,invariants,status,refines,any,when,where,with,then,begin,end,extends,sets,constants,axioms},
   sensitive=false,
  }

\lstdefinestyle{eventb-style}{
  language=eventb,
  basicstyle=\footnotesize\selectfont\rmfamily,
  keywordstyle=\fontseries{b}\selectfont\rmfamily,
  numbers=none
}

\begin{document}
\maketitle

\begin{abstract}
Ensuring the safety of autonomous vehicles (AVs) is the key requisite for their acceptance in society. This complexity is the core challenge in formally proving their safety conditions with AI-based black-box controllers and surrounding objects under various traffic scenarios. This paper describes our strategy and experience in modelling, deriving, and proving the safety conditions of AVs with the Event-B refinement mechanism to reduce complexity. Our case study targets the state-of-the-art model of goal-aware responsibility-sensitive safety to argue over interactions with surrounding vehicles. We also employ the Simplex architecture to involve advanced black-box AI controllers. Our experience has demonstrated that the refinement mechanism can be effectively used to gradually develop the complex system over scenario variations.
\end{abstract}


\section{Introduction}
The safety of automated vehicles has been attracting increased interest in society. In addition to the intensive effort of simulation-based testing, there is a key approach based on formal reasoning called responsibility-sensitive safety (RSS)~\cite{Shwartz2017RSS}. RSS defines the minimum rules that traffic participants should comply with for safety, i.e., no collisions. This rule-based approach has recently recently been extended to goal-aware RSS (GA-RSS) to deal with the goal-achievement, i.e., the driving goal of the ego-vehicle is eventually achieved such as pulling over upon emergency~\cite{Hasuo2022GARSS}. GARSS is effective for formally limiting liabilities, which is vital for AV manufacturers.

The challenge lies in deriving the necessary GARSS conditions and formally checking the compliance of the design of the ego vehicle over various scenarios under different environmental conditions. In addition, there is increasing demand to consider complex behaviours of black-box AI-based advanced controllers backed up with safety-ensured controllers, e.g., the Simplex architecture~\cite{Phan2017Simplex}.

Existing efforts have clarified the principles to derive and argue conditions that ego-vehicles should comply with in example scenarios. However, the engineering aspect has yet to be investigated. Specifically, we need a systematic modelling design that accepts the flexibility to mitigate the complexity in dealing with multiple aspects of scenario variations and architectural design.

To this end, we report our experience in modelling, deriving, and proving the safety conditions of autonomous vehicles (AVs). We follow the GA-RSS approach to define and derive the safety conditions to be checked with architectural design with black-box advanced controllers. We propose a strategy for using the refinement mechanism of Event-B~\cite{abrial2010modeling} to gradually argue the complex aspects including the scenario variations. Our experience has shown the potential of the refinement mechanism for the flexible design of models and proofs to mitigate the complexity in a gradual manner. To the best of our knowledge, this is the first attempt to focus on the model engineering aspect over scenario variations in the deductive approach for AV safety.

The rest of this paper is structured as follows: In \S~\ref{sec:preliminaries}, we describe the safety architecture, RSS, and Event-B. \S~\ref{sec:pullover} introduces GA-RSS and a case study example. We elaborate on our approach and its application to the case studies in \S~\ref{sec:ss4}--\ref{sec:ss3}. We discuss the approach in \S~\ref{sec:discussion} before concluding the paper in \S~\ref{sec:conclusion}.

\section{Preliminaries}\label{sec:preliminaries}

\subsection{Safety Architecture}\label{sec:safety-architecture}

Contemporary software systems often have black-box modules, such as machine learning modules, in which their safety is essentially difficult to verify.

\begin{figure}[tb]
  \centering
  \includegraphics[width=.8\linewidth]{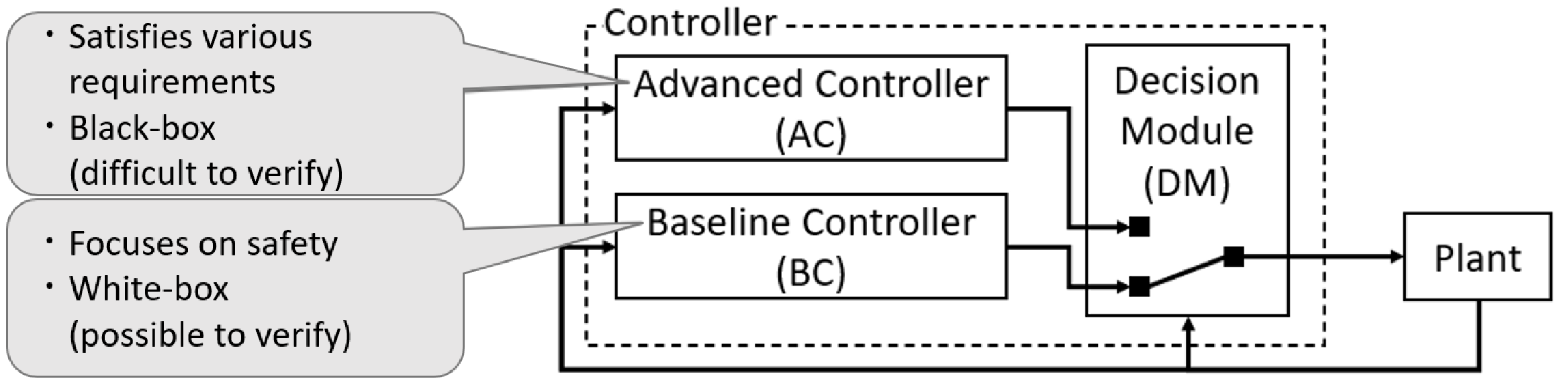}
  \caption{Component-based simplex architecture \cite{Phan2017Simplex}}
  \label{fig:simplex}
\end{figure}

A safety architecture, such as Simplex architecture (\figurename~\ref{fig:simplex}) \cite{Phan2017Simplex}, is a fundamental approach to guaranteeing the safety of such systems while benefitting from the high performance and functionality of black-box modules.
It models interactions between a controller and a plant.
The controller part has two different controllers: the baseline controller (BC), which is designed to force safe behaviour, and the advanced controller (AC), which aims at satisfying various requirements (e.g., comfort and progress) in addition to safety.
The decision module (DM) switches between the BC and AC in accordance with the state of the plant.
BC may fail to satisfy requirements other than safety, but it has a simple white-box behaviour enabling the safety to be easily verified.
In contrast, although AC usually gives better user experiences, guaranteeing its safety is difficult due to its complicated black-box behaviour.
For example, a typical BC for an AV may drive by following a predefined rule that is guaranteed to be safe in certain situations.
A typical AC, on the other hand, would be one that uses machine learning for motion planning.

\subsection{Responsibility-Sensitive Safety (RSS)}\label{sec:rss}
RSS is an approach to determining the safety of AVs by formal proof.
The core idea is to derive conditions that should be satisfied by the current state of the traffic participants such that safety, or no collisions, is  ensured in the future.

An RSS rule consists of an assertion $\precond$ called an RSS condition and a control strategy $\alpha$ called a proper response. They are defined for particular traffic scenarios. For example, a subject vehicle (SV), i.e., the ego vehicle, is following a preceding vehicle on a one-way road. We consider this preceding vehicle as the sole traffic participant called a principal other vehicle (POV). The SV must satisfy the RSS condition $\precond$ regarding the minimum relative distance from the POV. The distance is defined by considering the response time for braking and the distance necessary for the maximum comfortable braking to stop. The proper response $\alpha$ of the SV is to engage the maximum comfortable braking when the distance condition $\precond$ is about to be violated. The proof should show the RSS condition $\precond$ is preserved through the execution with the proper response $\alpha$.

In a general setting, RSS considers the SV and POV in the target scenario and determines the RSS condition and proper response. To prove the condition is preserved through the execution, a certain set of constraints must be satisfied by not only the SV but also all traffic participants (POVs), called RSS responsibility principles. Examples of the principles include ``do not cut in recklessly'' and ``be cautious in areas with limited visibility'', intuitively.

Our focus is not on the core responsibility principles of RSS but on the RSS-driven framework for proving safety of AVs. We are interested in the formal engineering aspect to model and verify scenario variations.

\subsection{Modelling and Proving in Event-B}
In this section, we describe the concepts of modelling and theorem proving in Event-B \cite{abrial2010modeling} that are used in our case study \footnote{For simplicity, we do not cover the ``full'' Event-B (described in \cite{abrial2010modeling}). For instance, our concrete machines inherit all variables and parameters from abstract machines, which is not necessary in general Event-B machines.}.

\subsubsection{Event-B Model Components.}
\begin{figure}[tb]
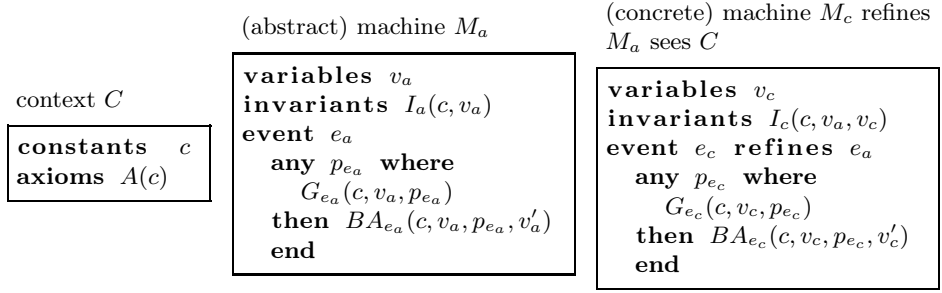


\centering

\begin{minipage}{.2\textwidth}
context $C$
   \begin{lstlisting}[language=eventb,mathescape]
constants  $c$
axioms $A(c)$
   \end{lstlisting}
\end{minipage}
\hspace{.025\textwidth}
\begin{minipage}{.35\textwidth}
(abstract) machine $M_a$
   \begin{lstlisting}[language=eventb,mathescape]
variables $v_a$
invariants $I_a(c, v_a)$
event $e_a$
  any $p_{e_a}$ where
    $G_{e_a}(c, v_a, p_{e_a})$
  then $BA_{e_a}(c, v_a, p_{e_a}, v_a')$
  end
   \end{lstlisting}
\end{minipage}
\hspace{.025\textwidth}
\begin{minipage}{.35\textwidth}
(concrete) machine $M_c$
refines $M_a$ sees $C$
   \begin{lstlisting}[language=eventb,mathescape]
variables $v_c$
invariants $I_c(c, v_a, v_c)$
event $e_c$ refines $e_a$
  any $p_{e_c}$ where
    $G_{e_c}(c, v_c, p_{e_c})$
  then $BA_{e_c}(c, v_c, p_{e_c}, v_c')$
  end
   \end{lstlisting}
\end{minipage}
  \caption{Structure of Event-B model components}
  \label{fig:event-b-model}
\end{figure}

Event-B models are structured as shown in \figurename~\ref{fig:event-b-model}.
The static aspects of the target system are specified as contexts, which consist of constants and their properties (axioms).
The dynamic aspects are specified as machines, which consist of variables, invariant predicates, and a set of events.
An event $e$ has parameters $p_e$, guard condition $G_e$, and before-after predicate $BA_e$ that explains the assignment performed in $e$ in terms of variables' current values $v$ and next values $v'$.
A significant feature of Event-B is a flexible refinement mechanism that enables declaring a machine $M_c$ as a refinement of another machine $M_a$.
Every event in $M_c$ should be seen as a refinement of events in $M_a$ (including the implicit \emph{skip} event).
$M_c$ does not need to inherit predicates of $M_a$, but those two machines should be compatible as described in the following.

\subsubsection{Proving Consistency of Models.}
Constructed models should be verified by discharging \emph{proof obligations} (POs) generated with predicates in the models.
Primary POs include the following:

\begin{itemize}
    \item \textbf{Invariant Preservation} (for an abstract machine): Invariant predicates are inductive ones, i.e., they must hold after every occurrence of events, given that they hold beforehand. Formally, invariant preservation by an event $e_a$ is: $A(c) \land I_a(c,v_a) \land G_{e_a}(c,v_a,p_{e_a}) \land BA_{e_a}(c,v_a,p_{e_a},v_a') \land \ldots \Longrightarrow I_a(c,v_a')$.
    \item \textbf{Invariant Preservation} (for concrete machines):  Formally, invariant preservation by an event $e_c$ is: $A(c) \land I_a(c,v_a) \land I_c(c,v_a,v_c) \land G_{e_c}(c,v_c,p_{e_c}) \land BA_{e_c}(c,v_c,p_{e_c},v_c') \land \ldots \Longrightarrow I_c(c,v_a',v_c')$.
    \item \textbf{Guard Strengthening}: For an event $e_c$ to be a refinement of an event $e_a$, the guard of $e_c$ must be stronger than that of $e_a$'s. Formally, guard strengthening of $e_c$ is: $A(c) \land I_c(c,v_a,v_c) \land  I_a(c, v_a) \land G_{e_c}(c,v_c,p_{e_c}) \land \ldots \Longrightarrow G_{e_a}(c,v_a,p_{e_a})$.
\end{itemize}

\section{Example: Goal-Aware RSS for Pull Over Scenario}
\label{sec:pullover}

\begin{figure}[tb]
    \centering
    \includegraphics[width=.9\linewidth]{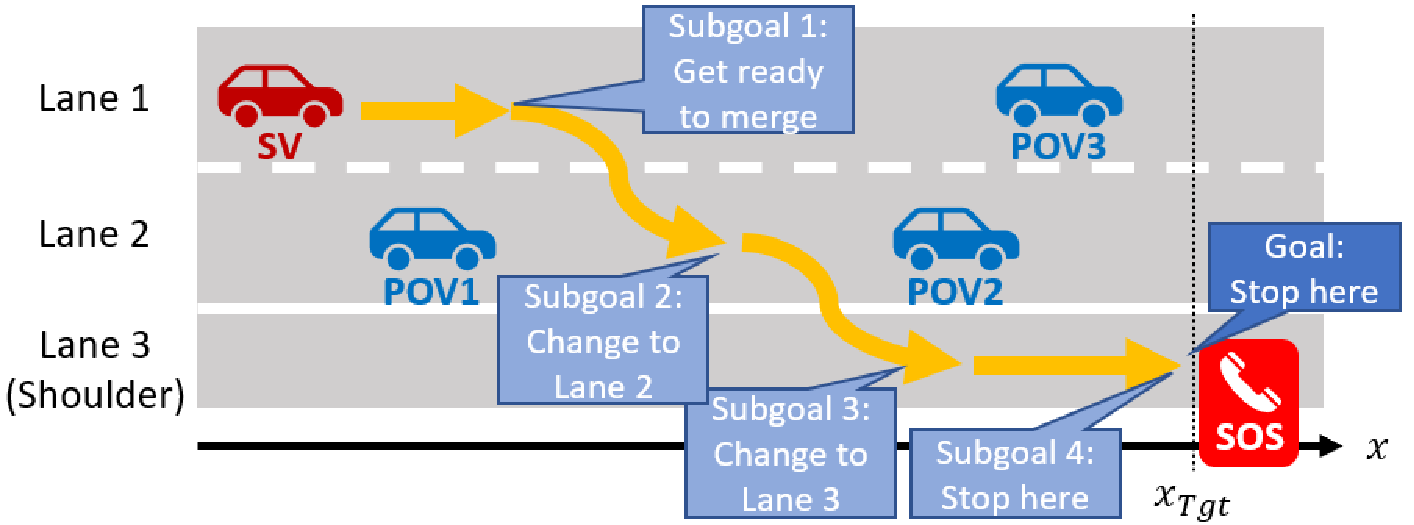}
    \caption{Pull over scenario \cite{Hasuo2022GARSS}}
    \label{fig:pull-over-scenario}
\end{figure}

Goal-aware RSS (GA-RSS) \cite{Hasuo2022GARSS} is an extension of RSS for dealing with complex scenarios that require planning over multiple manoeuvres to achieve particular goals.
For instance, consider the scenario shown in \figurename~\ref{fig:pull-over-scenario} (pull over scenario) \cite{Hasuo2022GARSS}: the SV needs to stop at a designated location ($\xtgt$) on the shoulder lane while keeping safe distances from POVs as required by RSS.
Following only the original RSS rules for avoiding collisions is necessary but not enough to achieve the goal.
The goal should be decomposed into several subgoals, such as (1) getting ready to merge between two POVs by changing the velocity, (2--3) changing lanes, and (4) stopping at $\xtgt$. Different proper responses are required for different subgoals as well.
However, for example, the SV can be trapped in Lane 1 if it is concerned about only the distance from the car ahead.

The workflow of GA-RSS is based on their extension of Floyd-Hoare logic.
Given a driving scenario $\scenario$ composed of the goal condition $\goal$ and safety condition $\safety$, the workflow is first used to decompose $\scenario$ into \emph{subscenarios} $\scenario_{1, \ldots, n}$ and identify the \emph{proper response} $\propresp_i$ for each subscenario $\scenario_{i}$.
\footnote{To be precise, with case distinctions, a tree of subscenarios is derived.}

Then, the \emph{precondition} $\precond_i$ for each subscenario is calculated as the precondition for establishing $\goal_i \land \precond_{i+1}$ while satisfying $\safety_{i}$, by performing $\propresp_i$.
Here, by seeing the (grand) goal of $\scenario$ as the postcondition of the final subscenario $\scenario_{n}$, the preconditions of all subscenarios are derived in a backward manner, \`a la Floyd-Hoare logic, and then integrated into the precondition of $\scenario$.

\begin{figure}[tb]
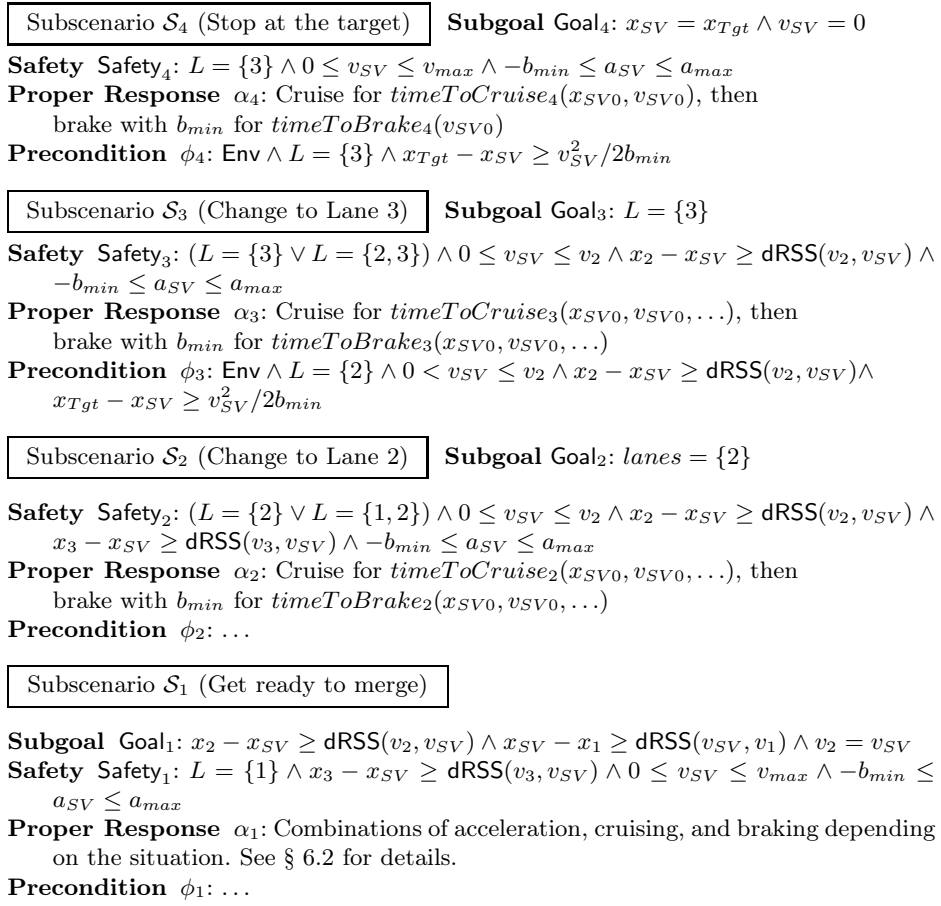

    \surroundrect{Subscenario $\scenario_{4}$ (Stop at the target)} \textbf{Subgoal} $\goal_4$: $\xsv = \xtgt \wedge \vsv = 0$
    \vspace{-5pt}
    \begin{description}
        \item[Safety] $\safety_4$: $\lanes = \{3\} \wedge 0 \leq \vsv \leq \vmax \wedge -\bmin \leq \asv \leq \amax$ 
        \item[Proper Response] $\propresp_4$: Cruise for $\timetocruise_4(\xsvinit, \vsvinit)$, then \\ brake with $\bmin$ for $\timetobrake_4(\vsvinit)$
        \item[Precondition] $\precond_4$: $\env \wedge \lanes = \{3\} \wedge \xtgt - \xsv \geq \vsv^2 / 2 \bmin$
    \end{description}

    \surroundrect{Subscenario $\scenario_{3}$ (Change to Lane 3)} \textbf{Subgoal} $\goal_3$: $\lanes = \{3\}$
    \vspace{-5pt}
    \begin{description}
        \item[Safety] $\safety_3$: $(\lanes = \{3\} \vee \lanes = \{2,3\}) \wedge 0 \leq \vsv \leq \vpov_2 \wedge \xpov_2 - \xsv \geq \drss(\vpov_2, \vsv) \wedge -\bmin \leq \asv \leq \amax$ 
        \item[Proper Response] $\propresp_3$: Cruise for $\timetocruise_3(\xsvinit, \vsvinit, \ldots)$, then \\ brake with $\bmin$ for $\timetobrake_3(\xsvinit, \vsvinit, \ldots)$
        \item[Precondition] $\precond_3$: $\env \wedge \lanes = \{2\} \wedge 0 < \vsv \leq \vpov_2 \wedge \xpov_2 - \xsv \geq \drss(\vpov_2, \vsv) \wedge$ \\ $\xtgt - \xsv \geq \vsv^2 / 2 \bmin$
    \end{description}

    \surroundrect{Subscenario $\scenario_{2}$ (Change to Lane 2)} \textbf{Subgoal} $\goal_2$: $lanes = \{2\}$
    \begin{description}
        \item[Safety] $\safety_2$: $(\lanes = \{2\} \vee \lanes = \{1,2\}) \wedge
        0 \leq \vsv \leq \vpov_2 \wedge
        \xpov_2 - \xsv \geq \drss(\vpov_2, \vsv) \wedge
        \xpov_3 - \xsv \geq \drss(\vpov_3, \vsv) \wedge
        -\bmin \leq \asv \leq \amax$ 
        \item[Proper Response] $\propresp_2$: Cruise for $\timetocruise_2(\xsvinit, \vsvinit, \ldots)$, then \\ brake with $\bmin$ for $\timetobrake_2(\xsvinit, \vsvinit, \ldots)$
        \item[Precondition] $\precond_2$: $\ldots$
    \end{description}

    \surroundrect{Subscenario $\scenario_{1}$ (Get ready to merge)}
    \begin{description}
        \item[Subgoal] $\goal_1$: $\xpov_2 - \xsv \geq \drss(\vpov_2,\vsv) \land \xsv - \xpov_1 \geq \drss(\vsv,\vpov_1) \land \vpov_2 = \vsv$ 
        \item[Safety] $\safety_1$: $\lanes = \{1\} \wedge
        \xpov_3 - \xsv \geq \drss(\vpov_3, \vsv) \wedge
        0 \leq \vsv \leq \vmax \wedge
        -\bmin \leq \asv \leq \amax$ 
        \item[Proper Response] $\propresp_1$: Combinations of acceleration, cruising, and braking depending on the situation. See \S~\ref{ssec:generality} for details.
        \item[Precondition] $\precond_1$: $\ldots$
    \end{description}
    \caption{Subscenarios of pull over scenario with proper response and precondition}
    \label{fig:pullover-subscenarios}
\end{figure}

For instance, \figurename~\ref{fig:pullover-subscenarios} shows the subgoals, safety conditions, proper responses, and preconditions of a subscenario chain (defined and derived in \cite{Hasuo2022GARSS}) where the SV goes between POV1 and POV2 and changes lanes.

Variables are as follows: $\xsv$ and $\xpov_{1,2,3}$ are the lateral positions of the SV and the three POVs; $\vsv$ and $\vpov_{1,2,3}$ are their lateral velocities; $\asv$ and $\apov_{1,2,3}$ are their lateral acceleration rates; $\lanes$ and $\lanes_{1,2,3}$ for set of lanes they are on.
Constants are as follows: $\xtgt$ is the position of the final goal position; $\vmin$ and $\vmax$ are the legal speed limits; $\bmin$ and $\bmax$ are the minimum (comfortable) and maximum (emergency) braking deceleration rates; $\amax$ is the maximum acceleration rate.

The condition of environment $\env$ is as follows:
\begin{align*}
    \env = &\bigwedge_{i=1,2,3}(\vmin \leq \vpov_i \leq \vmax \wedge \apov_i = 0) \\
           &\wedge \lanes_1 = \{2\} \wedge \lanes_2 = \{2\} \wedge \lanes_3 = \{1\} \wedge \xpov_2 > \xpov_1.
\end{align*}
This condition includes the assumption that POVs are supposed to run at constant velocity.

The RSS safety distance that the SV running at $\vsv$ should keep from the POV$i$ ahead running at $\vpov_i$ is defined as follows:
\begin{align}
    \drss(\vpov_i, \vsv) = \mathrm{max} \left( 0, \frac{\vsv^2}{2 \bmin} - \frac{\vpov_i^2}{2 \bmax} \right).
\end{align}

The times the SV should cruise, brake, or accelerate in subscenario $\scenario_{i}$ for proper response $\propresp_i$ are derived in the GA-RSS workflow \cite{Hasuo2022GARSS}. For instance,
\begin{align}
    \timetocruise_4(\xsvinit, \vsvinit) &= \frac{\xtgt - \xsvinit}{\vsvinit} - \frac{\vsvinit}{2 \bmin}, \label{eqn:timetocruisefour} \\
    \timetobrake_4(\vsvinit) &= \frac{\vsvinit}{\bmin}, \label{eqn:timetobrakefour}
\end{align}
where $\xsvinit$ and $\vsvinit$ are the position and velocity of the SV, respectively, when the switching occurs.

GA-RSS is designed to be integrated with the Simplex architecture.
The identified scenarios are used to construct the BC that performs the derived proper response $\propresp$ in the situation compatible with the scenario, and thus the BC is guaranteed to be safe and goal-achieving.
While the correctness of the DM is not covered with the method in \cite{Hasuo2022GARSS}, their experiment used their implementation of a Simplex-based controller, where the AC is black-box.

\textbf{Motivation of our case study.}
Even with the BC specifications identified with the GA-RSS workflow, a formal model of the whole Simplex architecture closer to the implementation is desired to construct safe and goal-achieving controllers of AVs.
Such models should at least take into account the behaviour of the DM and the monitor-decide-control loop (\figurename~\ref{fig:simplex}).

The challenge here is the model's \emph{complexity}; for example, in addition to DM-related elements, we need to take switching time delays into consideration.

To overcome this, we exploit the refinement mechanism of Event-B, which distributes the complexity of modelling and verification over multiple steps.

The rest of this paper discusses our case study, where we constructed and verified Event-B models of Simplex-based controllers for pull over subscenarios.

\section{Case Study 1: Modelling Subscenario $\scenario_4$}
\label{sec:ss4}

In this section, we introduce our modelling strategy, where elements of systems should be specified in each refinement step by using our model for subscenario $\scenario_4$ of the pull over scenario as an example.
We model the entire safety architecture and verify its safety in three refinement steps as follows:
\begin{description}
  \item[Machine $M_{4,0}$: Whole controller-level.]
        This is the most abstract machine.
        The properties of the whole controller's \\(AC+BC+DM) behaviour at every cycle are modelled.
        We focus on physical requirements that should be satisfied due to the controller's behaviour.
  \item[Machine $M_{4,1}$: Module-level.]
        This machine refines $M_{4,0}$.
        This machine is aware of the safety architecture; behavioural properties of AC, BC, and DM are specified separately.
        We checked that the switching by the DM satisfies the requirements in $M_{4,0}$ by proving the correctness of $M_{4,0}$--$M_{4,1}$ refinement.
  \item[Machine $M_{4,2}$: Manoeuvre-level.]
        This machine refines $M_{4,1}$.
        Details of the BC's behaviour (proper responses) are specified.
        By checking the correctness of $M_{4,1}$--$M_{4,2}$ refinement, we check that the proper responses satisfy the requirements.
\end{description}

\subsection{Machine $M_{4,0}$: Whole Controller-Level Behaviour}\label{sec:model0}

Machine $M_{4,0}$ is shown in \figurename~\ref{fig:model40}.
In this machine, we abstract away details of the controller and focus on the SV's position ($\xsv$) and velocity ($\vsv$) as the result of the controller's behaviour.

\begin{figure}[tb]

\centering

\begin{minipage}{.4\textwidth}
   \begin{lstlisting}[language=eventb,mathescape]
variables  $\xsv$, $\vsv$
   \end{lstlisting}
   \vspace{-8pt}
   \begin{lstlisting}[language=eventb,mathescape]
Event $\mathsf{initialisation}$
  any () where $\top$ then
    $\blabel{init\_sv}$: $(\xsv', \vsv')$ = $(\xsvinit, \vsvinit)$ end
   \end{lstlisting}
\end{minipage}
\hspace{.02\textwidth}
\begin{minipage}{.55\textwidth}
   \begin{lstlisting}[language=eventb,mathescape]
invariants
$\blabel{types}$: $\xsv \in \mathbb{R} \wedge \vsv \in \mathbb{R}$
$\blabel{no\_overrun}$: $0 \leq \xsv \leq \xtgt$
$\blabel{v\_regulated}$: $0 \leq \vsv \leq \vmax$
$\blabel{precond}$: $\xtgt - \xsv \geq \vsv^2 / 2 \bmin$
   \end{lstlisting}
\end{minipage}

%    \begin{lstlisting}[language=eventb,mathescape]
% variables  $\xsv$, $\vsv$
%    \end{lstlisting}

%    \begin{lstlisting}[language=eventb,mathescape]
% invariants
%  $\blabel{types}$: $\xsv \in \mathbb{R} \wedge \vsv \in \mathbb{R}$
%  $\blabel{no\_overrun}$: $0 \leq \xsv \leq \xtgt$
%  $\blabel{v\_regulated}$: $0 \leq \vsv \leq \vmax$
%  $\blabel{goal\_aware\_inv}$: $\xtgt - \xsv \geq \vsv^2 / 2 \bmin$
%  theorem $\blabel{goal}$: $\xsv = \xtgt \Longrightarrow \vsv = 0$
%    \end{lstlisting}

%    \begin{lstlisting}[language=eventb,mathescape]
% Event $\mathsf{initialisation}$
%   begin
%     $\blabel{init\_sv}$: $(\xsv', \vsv')$ = $(\xsvinit, \vsvinit)$
%   end
%    \end{lstlisting}

\vspace{-5pt}

   \begin{lstlisting}[language=eventb,mathescape]
Event $\mathsf{run}$
  any  $\xsvp$, $\vsvp$ where
    $\blabel{preserve\_no\_overrun}$: $0 \leq \xsvp \leq \xtgt$
    $\blabel{preserve\_v\_regulated}$: $0 \leq \vsvp \leq \vmax$
    $\blabel{preserve\_precond}$: $\xtgt - \xsvp \geq \vsvp^2 / 2 \bmin$
    $\blabel{x\_physical\_constr}$: $\xsv \leq \xsvp \leq \xsv + \int_{t=0}^{1} (\vsv + \amax t) dt$
    $\blabel{v\_physical\_constr}$: $\vsv - \int_{t=0}^{1} \bmax dt \leq \vsvp \leq \vsv + \int_{t=0}^{1} \amax dt$
  then
    $\blabel{update\_xv}$: $(\xsv', \vsv')$ = $(\xsvp, \vsvp)$ end
   \end{lstlisting}

  \caption{$M_{4,0}$: Abstract, whole controller-level machine for subscenario $\scenario_4$}
  \label{fig:model40}
\end{figure}

Invariant predicates \blabel{no\_overrun} and \blabel{v\_regulated} express basic requirements.

The precondition $\precond_4$ derived from the GA-RSS workflow is designed to be an invariant that the safety architecture should preserve; the DM enables using the AC while $\precond_4$ is \emph{robustly} satisfied, but it switches to the control using the BC once $\precond_4$ is \emph{about to be} violated.
Therefore, we specify $\precond_4$ as an invariant predicate (\blabel{precond}).

There is only a single non-initialisation event named $\mathsf{run}$.
It has parameters $\xsvp$ and $\vsvp$, which are specified as values of $\xsv$ and $\vsv$ at the next cycle (\blabel{update\_xv}).
The parameters are constrained by the guard predicates \blabel{preserve\_*} required for the event's invariant preservation and those for the constraints related to physics (\blabel{*\_physical\_constr}).
With these constraints as guard predicates of the event, we declare that every detailed behavioural description specified as events in concrete machines ($M_{4,1}$ and $M_{4,2}$) should satisfy the constraints.

The guard predicate \blabel{preserve\_precond} states that the controller \emph{somehow} produces the result (i.e., $\xsvp$ and $\vsvp$) such that \blabel{precond} is satisfied.
Indeed, the preservation of the precondition $\precond_4$ is trivial because:
\begin{align*}
    & (\xtgt - \xsvp \geq \vsvp^2 / 2 \bmin) \wedge \ldots \wedge ((\xsv', \vsv') = (\xsvp, \vsvp)) \\ 
    & \Longrightarrow \xtgt - \xsv' \geq \vsv'^2 / 2 \bmin.
\end{align*}
Note that \emph{how} the controller works to produce the invariant-satisfying result is not yet specified and deferred to concrete machines; how the DM prevents the AC from violating it is specified in machine $M_{4,1}$, and how the BC's behaviour (proper responses) satisfies it is specified in machine $M_{4,2}$.

\subsection{Machine $M_{4,1}$: Module-Level Behaviour}
\label{ssec:ss4m1}

In machine $M_{4,1}$ (\figurename~\ref{fig:model41}), which refines $M_{4,0}$, we focus on the requirements on white-box modules of the architecture, namely the BC and DM, particularly the condition for switching; through the proof attempt, we \emph{derived} the switching condition such that the precondition is always satisfied.
Note that we assume that the AC's behaviour is arbitrary as long as it satisfies $\mathsf{run}$'s guard.
Details of the BC's behaviour that should be specified using the time spent for each manoeuvre are introduced in machine $M_{4,2}$.

\begin{figure}[tb]

\centering
   \begin{lstlisting}[language=eventb,mathescape]
variables  $\xsv$, $\vsv$, $\ctrl$, $\vsvbcinit$
   \end{lstlisting}

   \vspace{-8pt}
   \begin{lstlisting}[language=eventb,mathescape]
invariants
$\blabel{types}$: $\ctrl \in \{\ac, \bc\} \wedge \vsvbcinit \in \mathbb{R}$
$\blabel{vsvbcinit\_regulated}$: $0 \leq \vsvbcinit \leq \vmax$
$\blabel{bc\_no\_accel}$: $\ctrl = \bc \Longrightarrow \vsv \leq \vsvbcinit$
$\blabel{switching}$: $\ctrl = \ac \Longrightarrow \gainv_4(\xsv + \int_{t=0}^{1} (\vsv + \amax t) dt, \; \vsv + \int_{t=0}^{1} \amax dt)$
   \end{lstlisting}

%    \begin{lstlisting}[language=eventb,mathescape]
% Event $\mathsf{initialisation}$ ...
%    \end{lstlisting}

%    \begin{lstlisting}[language=eventb,mathescape]
% Event $\mathsf{initialisation}$
%   begin
%     $\blabel{init\_sv}$: $(\xsv', \vsv')$ = $(\xsvinit, \vsvinit)$
%     $\blabel{init\_ctrl}$: $\ctrl' = \ac \Longleftrightarrow \gainv(\xsvinit + \int_{t=0}^{1} (\vsvinit + \amax t) dt, \; \vsvinit + \int_{t=0}^{1} \amax dt)$
%     $\blabel{init\_vsvbcinit}$: $\vsvbcinit' = \vsvinit$
%   end
%    \end{lstlisting}

%    \vspace{-8pt}
%    \begin{lstlisting}[language=eventb,mathescape]
% Event $\mathsf{AC \rightarrow AC}$ refines $\mathsf{run}$ 
%   any  $\xsvp$, $\vsvp$ where
%     $\ldots$ (guard predicates of $\mathsf{run}$ except $\blabel{preserve\_goal\_aware\_inv}$) $\ldots$
%     $\blabel{AC\_operating}$: $\ctrl = \ac$
%     $\blabel{surely\_safe\_next}$: $\gainv(\xsv + \int_{t=0}^{2} (\vsv + \amax) dt , \; \vsv + \int_{t=0}^{2} \amax dt)$
%   then
  %   $\ldots$ (actions of $\mathsf{run}$) $\ldots$
  % end
  %  \end{lstlisting}

   \vspace{-8pt}
   \begin{lstlisting}[language=eventb,mathescape]
Event $\mathsf{AC \rightarrow BC}$ refines $\mathsf{run}$ 
  any  $\xsvp$, $\vsvp$ where
    $\ldots$ (guard predicates of $\mathsf{run}$ except $\blabel{preserve\_precond}$) $\ldots$
    $\blabel{AC\_operating}$: $\ctrl = \ac$
    $\blabel{maybe\_unsafe\_next}$: $\neg \gainv_4(\xsv + \int_{t=0}^{2} (\vsv + \amax) dt , \; \vsv + \int_{t=0}^{2} \amax dt)$
  then
    $\ldots$ (actions of $\mathsf{run}$) $\ldots$
    $\blabel{switch\_to\_bc}$: $\ctrl' = \bc$
    $\blabel{vsvbcinit\_update}$: $\vsvbcinit' = \vsvp$ end
   \end{lstlisting}

%    \vspace{-8pt}
%    \begin{lstlisting}[language=eventb,mathescape]
% Event $\mathsf{BC \rightarrow BC}$ refines $\mathsf{run}$ 
%   any  $\xsvp$, $\vsvp$ where
%     $\ldots$ (guard predicates of $\mathsf{run}$) $\ldots$
%     $\blabel{BC\_operating}$: $\ctrl = \bc$
%     $\blabel{no\_acceleration}$: $\vsvp \leq \vsvbcinit$
%     $\blabel{maybe\_unsafe\_next}$: $\neg \gainv(\xsv + \int_{t=0}^{2} (\vsv + \amax) dt , \; \vsv + \int_{t=0}^{2} \amax dt)$
%   then
%     $\ldots$ (actions of $\mathsf{run}$) $\ldots$
%   end
%    \end{lstlisting}

   \vspace{-8pt}
   \begin{lstlisting}[language=eventb,mathescape]
Event $\mathsf{BC \rightarrow AC}$ refines $\mathsf{run}$ 
  any  $\xsvp$, $\vsvp$ where
    $\ldots$ (guard predicates of $\mathsf{run}$) $\ldots$
    $\blabel{BC\_operating}$: $\ctrl = \bc$
    $\blabel{no\_acceleration}$: $\vsvp \leq \vsvbcinit$
    $\blabel{surely\_safe\_next}$: $\gainv_4(\xsv + \int_{t=0}^{2} (\vsv + \amax) dt , \; \vsv + \int_{t=0}^{2} \amax dt)$
  then
    $\ldots$ (actions of $\mathsf{run}$) $\ldots$
    $\blabel{switch\_to\_ac}$: $\ctrl' = \ac$ end
   \end{lstlisting}

  \caption{(A part of) $M_{4,1}$: Intermediate, module-level machine for subscenario $\scenario_4$}
  \label{fig:model41}
\end{figure}

\label{ssec:machine41}

There are two new variables: $\ctrl$, for the currently active controller, and $\vsvbcinit$, which stores the velocity at the time when switching to the BC occurs.

Invariant predicates are in regard to the requirements on the BC and DM. \\
\blabel{vsvbcinit\_regulated} requests that $\vsvbcinit$ should not exceed $\vmax$ like $\vsv$, and \blabel{bc\_no\_accel} expresses that the BC does not accelerate in the proper response.
\blabel{switching} states that if the AC is active, then the SV will be goal-achieving and safe after a cycle even if the SV accelerated with the maximum rate $\amax$.
The contraposition of \blabel{switching} means that the BC is used if the precondition $\precond_4$ may be violated at the next cycle.

There are four events for cases of switching: $\mathsf{AC \rightarrow AC}$, $\mathsf{AC \rightarrow BC}$, $\mathsf{BC \rightarrow BC}$, and $\mathsf{BC \rightarrow AC}$.
They all refine the $\mathsf{run}$ event of the previous machine $M_{4,0}$.
For instance, $\mathsf{AC \rightarrow BC}$ is for the case where the current controller is the AC (\blabel{AC\_operating}) and \blabel{switching} can be violated after the event (\blabel{maybe\_unsafe\_next}; note that the integrals are from $t=0$ to $2$ to look ahead for two cycles).
Note that, however, \blabel{switching} is guaranteed to hold before the event since it is an invariant predicate. In addition to actions of $\mathsf{run}$, the controller is switched to the BC (\blabel{switch\_to\_bc}) and $\vsvbcinit$ is updated (\blabel{vsvbcinit\_update}).
On the other hand, $\mathsf{BC \rightarrow AC}$ is the case where the controller is switched from the BC to AC because the invariant \blabel{switching} will be satisfied after the occurrence of the event (\blabel{surely\_safe\_next}).

The main POs are as follows:
\\
    1. \textbf{Do the events $\mathsf{AC \rightarrow *}$ preserve the invariant \blabel{precond}?}
    This corresponds to the guard strengthening PO of $\mathsf{AC \rightarrow *}$. The intuition of the proof is because the AC is operating only if the precondition is guaranteed to hold after two cycles (\blabel{surely\_safe\_next}), and it is guaranteed to hold after one cycle as well.
    \\
    2. \textbf{Do events $\mathsf{* \rightarrow AC}$ preserve the invariant \blabel{switching}?} 
    It is preserved because the AC will be used only if \blabel{surely\_safe\_next} holds at the current state.
    In fact, we \emph{derived} the switching condition \blabel{surely\_safe\_next} through the attempt to discharge this PO.

\subsection{Machine $M_{4,2}$: Manoeuvre-Level Behaviour}
In machine $M_{4,2}$ (\figurename~\ref{fig:model42-1}), which refines $M_{4,1}$, we focus on the details of the behaviour with the notion of time to spend on each manoeuvre to verify that the BC's behaviour satisfies the requirements specified in machines $M_{4,0}$ and $M_{4,1}$.

\begin{figure}[tb]

\centering
   \begin{lstlisting}[language=eventb,mathescape]
variables  $\xsv$, $\vsv$, $\ctrl$, $\vsvbcinit$, $\tcruisebc$, $\tbrakebc$
   \end{lstlisting}

   \vspace{-8pt}
   \begin{lstlisting}[language=eventb,mathescape]
invariants
$\blabel{types}$: $\tcruisebc \in \mathbb{R}_{\geq 0} \wedge \tbrakebc \in \mathbb{R}_{\geq 0}$
$\blabel{cruise\_before\_brake}$: $0 < \tcruisebc \Longrightarrow 0 < \tbrakebc$
$\blabel{v\_in\_BC}$: $\ctrl = \bc \Longrightarrow \vsv = \tbrakebc \cdot \bmin$
$\blabel{v\_in\_BC\_cruise}$: $(\ctrl = \bc \wedge 0 < \tcruisebc) \Longrightarrow \vsv = \vsvbcinit$
$\blabel{x\_in\_BC}$: $(\ctrl = \bc \wedge \vsv \neq 0)$ 
$\Longrightarrow \xtgt - \xsv = \int_{t=0}^{\tcruisebc} \vsv dt + \int_{t=0}^{\tbrakebc} (\vsv - \bmin t) dt$
   \end{lstlisting}

   \vspace{-8pt}
   \begin{lstlisting}[language=eventb,mathescape]
axioms $\blabel{t\_bccruise\_def}$: $\timetocruise_4(x,v) = (2 \bmin (\xtgt - x) - v^2) / (2 \bmin v)$
   \end{lstlisting}

%    \vspace{-8pt}
%    \begin{lstlisting}[language=eventb,mathescape]
% Event $\mathsf{initialisation}$ ...
%    \end{lstlisting}

   % \vspace{-8pt}
%    \begin{lstlisting}[language=eventb,mathescape]
% Event $\mathsf{initialisation}$
%   begin
%     $\ldots$ (initialisation actions for other $\mathrm{variables}$) $\ldots$
%     $\blabel{init\_tcruisebc}$: $\tcruisebc' = \timetocruise_4(\xsvinit, \vsvinit)$
%     $\blabel{init\_tbrakebc}$: $\tbrakebc' = \vsvinit / \bmin$
%   end
%    \end{lstlisting}

%    \vspace{-8pt}
%    \begin{lstlisting}[language=eventb,mathescape]
% Event $\mathsf{AC \rightarrow AC}$ refines $\mathsf{AC \rightarrow AC}$ 
%   $\ldots$ (the same as Model 1) $\ldots$
%    \end{lstlisting}

   \vspace{-8pt}
   \begin{lstlisting}[language=eventb,mathescape]
Event $\mathsf{AC\_run \rightarrow BC}$ refines $\mathsf{AC \rightarrow BC}$ 
  any  $\xsvp$, $\vsvp$ where
    $\ldots$ (guard predicates of $\mathsf{AC \rightarrow BC}$) $\ldots$ 
    $\blabel{will\_run\_more}$: $0 < \vsvp$
  then
    $\ldots$ (actions of $\mathsf{AC \rightarrow BC}$) $\ldots$
    $\blabel{update\_tcruisebc}$: $\tcruisebc' = \timetocruise_4(\xsvp, \vsvp)$
    $\blabel{update\_tbrakebc}$: $\tbrakebc' = \timetobrake_4(\vsvp)$
  end
   \end{lstlisting}

%    \vspace{-8pt}
%    \begin{lstlisting}[language=eventb,mathescape]
% Event $\mathsf{AC\_stop \rightarrow BC}$ refines $\mathsf{AC \rightarrow BC}$ $\ldots$
%    \end{lstlisting}

   \vspace{-8pt}
   \begin{lstlisting}[language=eventb,mathescape]
Event $\mathsf{BC\_cruise \rightarrow AC}$ refines $\mathsf{BC \rightarrow AC}$ 
  any  $\xsvp$, $\vsvp$ where
    $\blabel{BC\_operating}$: $\ctrl = \bc$
    $\blabel{surely\_safe\_next}$: $\gainv_4(\xsv + \int_{t=0}^{2} (\vsv + \amax) dt , \; \vsv + \int_{t=0}^{2} \amax dt)$
    $\blabel{will\_cruise\_more}$: $1 \leq \tcruisebc$
    $\blabel{cruise\_xv}$: $\xsvp = \xsv + \int_{t=0}^{1} \vsv dt \wedge \vsvp = \vsv + \int_{t=0}^{1} 0 dt$
  then
    $\ldots$ (actions of $\mathsf{BC \rightarrow AC}$) $\ldots$
    $\blabel{tcruise\_pass}$: $\tcruisebc' = \tcruisebc - 1$
  end
   \end{lstlisting}

  \caption{(A part of) $M_{4,2}$: Concrete, manoeuvre-level machine for subscenario $\scenario_4$}
  \label{fig:model42-1}
\end{figure}

Two new variables about the remaining time for cruising ($\tcruisebc$) and braking ($\tbrakebc$) are introduced.
The unit of time here is the cycle, e.g., the value of $\tcruisebc$ is the number of the controller's cycles spent for cruising.

\begin{figure}[tb]
    \centering
    \begin{minipage}{.45\textwidth}
     \includegraphics[width=0.99\linewidth]{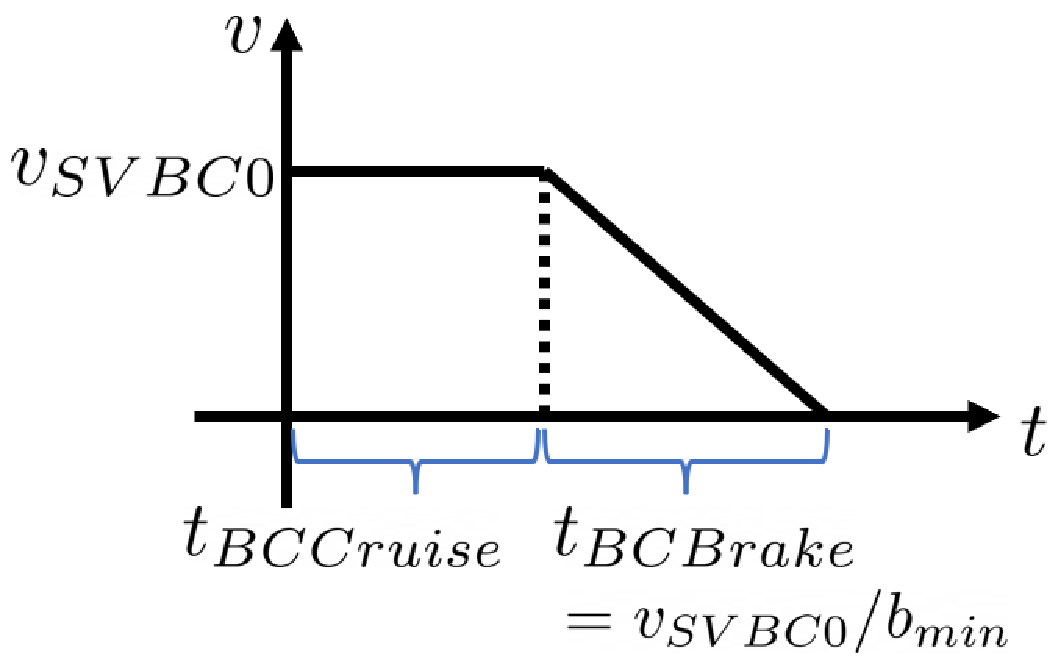}
     \caption{Proper response $\propresp_4$}
     \label{fig:proper-response}
    \end{minipage}
    \qquad
    \begin{minipage}{.45\textwidth}
     \includegraphics[width=.99\linewidth]{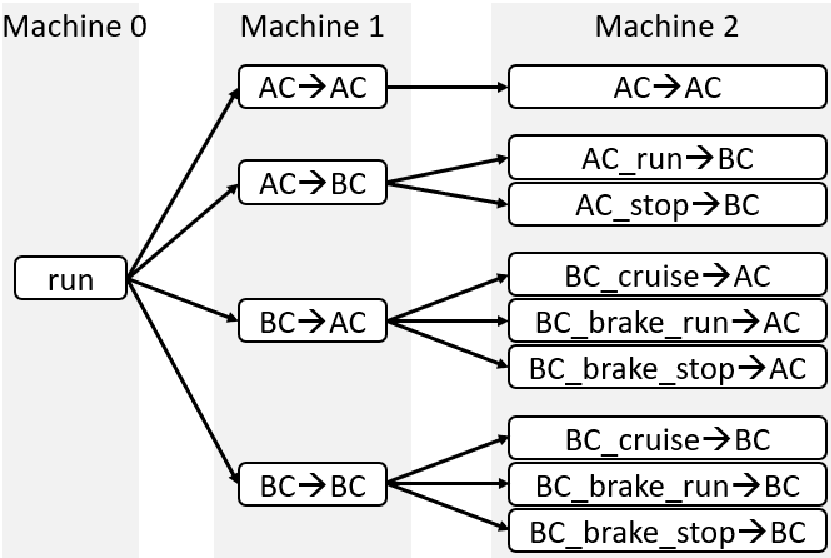}
     \caption{Event refinement relationship}
     \label{fig:events-refinement}
    \end{minipage}
\end{figure}

Invariant predicates are in regard to the detailed properties of the BC's behaviour: \blabel{cruise\_before\_brake} expresses that the proper response $\propresp_4$ is cruising and then braking, and \blabel{*\_in\_BC*} states that the velocity and position should follow the proper response $\propresp_4$ as shown in \figurename~\ref{fig:proper-response}.

Events of $M_{4,2}$ refine those of $M_{4,1}$ as shown in \figurename~\ref{fig:events-refinement}.

Three events that refine $\mathsf{AC \rightarrow *}$ are mostly the same as $M_{4,1}$, but events regarding switching to the BC (such as $\mathsf{AC\_run \rightarrow BC}$) are extended with actions of calculating $\tcruisebc$ and $\tbrakebc$ as equations \ref{eqn:timetocruisefour} and \ref{eqn:timetobrakefour} (derived in the GA-RSS workflow) because the BC should calculate them every time it get activated.

Unlike events that refine $\mathsf{AC \rightarrow *}$, six events that refine $\mathsf{BC \rightarrow *}$ do not inherit all of the guard predicates and actions of corresponding events in machine $M_{4,1}$.
For example, the differences between the event $\mathsf{BC\_cruise \rightarrow AC}$ in $M_{4,2}$ and the corresponding event $\mathsf{BC \rightarrow AC}$ in $M_{4,1}$ is as shown in \figurename~\ref{fig:bccruise-diff}.
The removed guard predicates (lines with red background) are requirements on the values of the SV's position and velocity after the occurrence of the event ($\xsvp$ and $\vsvp$), while introduced guard predicates (lines with green background) include the concrete behaviour of the BC (\blabel{cruise\_xv}), namely running with the constant velocity.
By changing events in this way and checking that the guard of $\mathsf{BC\_cruise \rightarrow AC}$ is stronger than that of $\mathsf{BC \rightarrow AC}$, we can verify that the BC's concrete behaviour satisfies the requirements specified in machines $M_{4,0}$ and $M_{4,1}$.

\begin{figure}[tb]

 \begin{center}

\begin{lstlisting}[language=eventb,mathescape,linebackgroundcolor={\ifcase \value{lstnumber} \or \color{minuscolor} \or \color{pluscolor} \or\or \color{minuscolor} \or \color{minuscolor} \or \color{minuscolor} \or \color{minuscolor} \or \color{minuscolor} \or \or \color{minuscolor} \or \or \color{pluscolor} \or \color{pluscolor} \or \or \or \or \color{pluscolor}\fi}]
-Event $\mathsf{BC \rightarrow AC}$ refines $\mathsf{run}$ 
+Event $\mathsf{BC\_cruise \rightarrow AC}$ refines $\mathsf{BC \rightarrow AC}$ 
  any  $\xsvp$, $\vsvp$ where
-   $\blabel{preserve\_no\_overrun}$: $0 \leq \xsvp \leq \xtgt$
-   $\blabel{preserve\_v\_regulated}$: $0 \leq \vsvp \leq \vmax$
-   $\blabel{preserve\_precond}$: $\xtgt - \xsvp \geq \vsvp^2 / 2 \bmin$
-   $\blabel{x\_physical\_constr}$: $\xsv \leq \xsvp \leq \xsv + \int_{t=0}^{1} (\vsv + \amax t) dt$
-   $\blabel{v\_physical\_constr}$: $\vsv - \int_{t=0}^{1} \bmax dt \leq \vsvp \leq \vsv + \int_{t=0}^{1} \amax dt$
    $\blabel{BC\_operating}$: $\ctrl = \bc$
-   $\blabel{no\_acceleration}$: $\vsvp \leq \vsvbcinit$
    $\blabel{surely\_safe\_next}$: $\gainv_4(\xsv + \int_{t=0}^{2} (\vsv + \amax) dt , \; \vsv + \int_{t=0}^{2} \amax dt)$
+   $\blabel{will\_cruise\_more}$: $1 \leq \tcruisebc$
+   $\blabel{cruise\_xv}$: $\xsvp = \xsv + \int_{t=0}^{1} \vsv dt \wedge \vsvp = \vsv + \int_{t=0}^{1} 0 dt$
  then
    $\blabel{update\_xv}$: $(\xsv', \vsv')$ = $(\xsvp, \vsvp)$
    $\blabel{switch\_to\_ac}$: $\ctrl' = \ac$
+   $\blabel{tcruise\_pass}$: $\tcruisebc' = \tcruisebc - 1$
  end
\end{lstlisting}

\end{center}
\caption{Differences between $\mathsf{BC \rightarrow AC}$ (in $M_{4,1}$) and $\mathsf{BC\_cruise \rightarrow AC}$ (in $M_{4,2}$)}
\label{fig:bccruise-diff}
\end{figure}

In addition to the consistency between the BC's concrete behaviour specified in $M_{4,2}$ and requirements on the BC specified in $M_{4,1}$, we checked that events $\mathsf{* \rightarrow BC}$ and $\mathsf{BC \rightarrow *}$ preserve the invariant.

\section{Case Study 2: Modelling Subscenario $\scenario_3$}
\label{sec:ss3}
In this section, we use subscenario $\scenario_3$ to demonstrate how our modelling strategy (\S~\ref{sec:ss4}) is applicable to other subscenarios.
subscenario $\scenario_3$ has new aspects; the SV is changing lanes and the leading vehicle POV2.

\subsection{Machine $M_{3,0}$: Whole Controller-Level Behaviour}

Following machine $M_{4,0}$ of subscenario $\scenario_4$, we focus only on the physical results of the controller behaviour.

POV2's variable position ($\xLead$) and constant velocity ($\vLead$) are used in addition to SV's position and velocity.

As the SV is changing lanes, we assume that this action will be done in an exact amount of time modelled as a constant $\tLC$ (the time for lane changing), and therefore we introduce another variable $\tLCe$ (the time for lane changing elapsed) so that when the time elapsed reaches $\tLC$, the SV should have finished switching lanes and the subscenario is over. We modelled lanes in this style instead of introducing another physical coordinate for simplicity.

A new invariant predicate \blabel{no\_overtime} regarding the time limit of this subscenario is also introduced as a replacement for \blabel{no\_overrun} of subscenario $\scenario_4$.
The corresponding guard predicates of the event $\mathsf{run}$ are specified so that no event can occur once the lane switching is over.
\begin{center}
\begin{tabular}{|c|}
  \hline
  \blabel{no\_overtime}: $\tLCe \leq \tLC$ \\
  \hline
\end{tabular}
\end{center}
The precondition for subscenario $\scenario_3$ ($\precond_3$ derived in \cite{Hasuo2022GARSS}) takes into consideration the RSS safety distance between the SV and the leading vehicle POV2.
\begin{center}
\begin{tabular}{|cl|}
  \hline
  \blabel{precond}: & $\xtgt - \xsv \geq \vsv^2 / 2 \bmin    \land    \xsv < \xLead$ \\
  & $\land    2(\xsv - \xLead) + \frac{\vsv^2}{\bmin} \leq \frac{\vLead^2}{\bmax}$\\
  \hline
\end{tabular}
\end{center}

As in subscenario $\scenario_4$, the $\mathsf{run}$ event has guard predicates to preserve invariant predicates.
The event also has new actions for updating $\xLead$ and $\tLCe$:
\begin{center}
\begin{tabular}{|cl|}
  \hline
  \blabel{update\_xLead}: & $\xLead' = \xLead + \int_{t=0}^{1} (\vLead t) dt$ \\
  \blabel{update\_xLCe}: & $\tLCe' = \mathrm{min}(\tLC, \tLCe + 1)$\\
  \hline
\end{tabular}
\end{center}

\subsection{Machine $M_{3,1}$: Module-Level Behaviour}

This machine is also similar to $M_{4,1}$, but the invariant \blabel{switching} and guard predicates \blabel{surely\_safe\_next} (and its negation \blabel{maybe\_unsafe\_next}) take into account the distance between the SV and POV2.
\begin{center}
\begin{tabular}{|cl|}
  \hline
  \blabel{switching}: & $\ctrl = \ac \Longrightarrow \gainv_3(\xsv + \int_{t=0}^{1} (\vsv + \amax t) dt, \;$ \\
                        & $\vsv + \int_{t=0}^{1} \amax dt,  \; \xLead + \int_{t=0}^{1} (\vLead t) dt ,  \; \vLead)$ \\
  \hline
\end{tabular}
\end{center}
\vspace{-15pt}
\begin{center}
\begin{tabular}{|cl|}
  \hline
  \blabel{surely\_safe\_next}: & $\gainv_3(\xsv + \int_{t=0}^{2} (\vsv + \amax t) dt, \; \vsv + \int_{t=0}^{2} \amax dt,  \;$ \\
                            & $ \xLead + \int_{t=0}^{2} (\vLead t) dt ,  \; \vLead)$ \\
  \hline
\end{tabular}
\end{center}

As subscenario $\scenario_4$ (\S~\ref{ssec:ss4m1}), the POs are in regard to the preservations of invariants \blabel{precond} and \blabel{switching}.

\subsection{Machine $M_{3,2}$: Manoeuvre-Level Behaviour}

Compared with $M_{4,2}$ for subscenario $\scenario_4$, there are two major differences: when switching to the BC, the calculation of $\tcruisebc$ and $\tbrakebc$ (derived in \cite{Hasuo2022GARSS}) is different because the velocity of the SV should not be zero by the end of the subscenario $\scenario_3$ but only low enough to satisfy the goal invariant.

\begin{center}
\begin{tabular}{|cl|}
  \hline
  \blabel{tBrake\_update}: & $\tbrakebc' = (\tLC - \tLCe) + \frac{\vsvp}{2.\bmin} + \frac{\xsvp - \xtgt}{\vsvp}$\\
  \hline
\end{tabular}
\end{center}

The six events that refine \textbf{$\mathsf{BC\_* \rightarrow *}$} have to satisfy machine $M_{3,0}$'s \blabel{precond} that now includes the safety distance to the leading vehicle POV2.

The POs in regard to this invariant were discharged in the following way:
\begin{enumerate}
    \item \textbf{$\mathsf{BC\_* \rightarrow BC}$} The idea behind this proof is that BC's proper response does not include accelerating and the leading vehicle's velocity is constant, so the distance between these two may only increase.
    \item \textbf{$\mathsf{BC\_* \rightarrow AC}$}
    The guard predicate \blabel{surely\_safe\_next} states that the invariant will be satisfied in two cycles without having to break in the next cycle because the controller will be in the AC.
\end{enumerate}

\section{Discussion}
\label{sec:discussion}
\subsection{Model Engineering}
In the case studies, we have used the refinement mechanism of Event-B to gradually model and verify the different aspects. Specifically, we separated the argument over the definition of safe and goal-achieving behaviour, architecture for switching behaviours, and concrete behaviour design. The refinement mechanism limits the complexity of modelling and proof in each step, which was essential in handling the increasing complexity in proving continuous properties.

We did not directly reuse the models between subscenarios, e.g., sharing the abstract steps between subscenarios. This is our explicit choice as the key safety properties and involved variables for the POVs are unique to each subscenario. We instead used the common refinement strategy as well as the model representations. We believe this experience enables us to demonstrate the know-how for scenarios other than the pull over scenario. The generality of the approach is further discussed in the following.

\subsection{Generality of Approach}
\label{ssec:generality}
We have described how the same refinement strategy can deal with subscenarios $\scenario_3$ and $\scenario_4$. We describe how the other subscenarios can be modelled as well as the omitted aspect of perception errors.

\subsubsection{Subscenario $\scenario_2$.}
The machines for subscenario $\scenario_2$ are similar to that for subscenario $\scenario_3$. The main difference between them is the presence of a leading vehicle in the next lane in subscenario $\scenario_2$ while there is none in subscenario $\scenario_3$.

\subsubsection{Subscenario $\scenario_1$.}
In this subscenario, the SV needs to prepare to switch lanes and merge into the next lane. There are three POVs to take into account: one ahead of the SV in the current lane (POV3) and two others in the next lane (POV 1 and 2). This subscenario thus involves multiple (in this case, four) proper responses: an example is accelerating to pass POV1 in the next lane, and another example is decelerating to match the velocity of POV2 in the next lane. 
To handle multiple proper responses in a unified manner, we modelled them as a sequence of proper responses with variable durations as follows: (1) Accelerate for $t_{BCAccel}$ (2) Cruise for $t_{BCCruise}$ (3) Brake for $t_{BCBrake}$.
Moreover, we needed to take into account different precondition for each proper response. Therefore, we introduced a variable to record which proper response was taken the last time the BC got activated.

\subsubsection{Perceptual Uncertainty}
Another aspect not included in the case studies is perceptual uncertainty or the possibility of errors in sensing. A basic approach to this issue would be adding safety margins to the behaviour of the controller. For instance, introducing a variable $\widehat{\xtgt}$ for the perceived value of target location ($\xtgt$) and discussing assumptions on the difference between $\xtgt$ and $\widehat{\xtgt}$ enables us to derive the appropriate amount of the safety margin for this uncertainty.

\subsection{Using Event-B for Modelling and Proving}
Features of Event-B and its modelling environment Rodin \cite{Abrial2010Rodin} were useful for modelling and proving the safety architecture for GA-RSS.
Rodin generated POs and helped interactive proof of them.
The refinement mechanism of Event-B was effective for distributing the complexity of modelling and proving over multiple steps.
In addition, as we discussed in \S \ref{ssec:machine41}, we derived the correct behaviour of DM from generated POs.

Our contributions in this paper, namely strategies of modelling and refinement, provide a guide to the effective use of Event-B's features for the rigorous and systematic construction of controllers for different subscenarios.

On the other hand, although Rodin has proof tactics and provers for automatically discharging POs, we had to manually discharge all POs.
It is because we needed an extension of Event-B language \cite{Butler2013} to use real numbers in models, and Rodin's current automatic proof functionalities are not strong when the language is extended.
However, we expect that this problem will be solved; for instance, there are studies aiming at assisting automatic proof of hybrid systems by bridging Rodin with external solvers \cite{9763726}.

\subsection{Related Work}
RSS was originally proposed as the formal approach for AVs, but the paper did not include any machine-processible models~\cite{Shwartz2017RSS}.
The work on GA-RSS extended the framework of RSS with formal specifications and partial calculations supported by Mathematica~\cite{Hasuo2022GARSS}.
Other studies only used the resulting RSS conditions, for example, encoding them in signal temporal logic for runtime verification~\cite{Hekmatnejad2019RSSMonitoring}. 
To the best of our knowledge, this is the first attempt to make use of formal modelling for the RSS scheme. The study in \cite{Roohi2018RSSchecking} demonstrated the difficulty in checking RSS properties with automated ``one button'' tools for reachability analysis and model checking.

Other formal attempts for AVs include proofs with the Isabelle/HOL prover~\cite{Rizaldi2018AVbyIsabelle} with support of MATLAB. The focus was on the detailed computation including floating-point errors while the driving behaviour was rather simple; avoidance of one static object with a white-box controller.

Verification over RSS is intrinsically hybrid, i.e., including continuous aspects such as velocity and distance. Proofs over hybrid models have been actively investigated in the Hoare-style reasoning, not only for Event-B but also in other formalisms such as KeYmaera X~\cite{Fulton2015KeYmaeraX}. Our case study did not focus on the continuous aspects and used rather simple theories for handling real arithmetic. Our future work includes the use of more sophisticated support for discharging the proof obligations. It is notable that refining continuous models in the physics world into discrete software controllers has been actively investigated for Event-B, e.g., ~\cite{Dupont2021EventBHybridation}. Models obtained in our approach can be further refined with such techniques into concrete designs of discrete software controllers.

Guidelines with a focus on refinement strategies have been considered useful for Event-B as reusable know-how for specific types of systems~\cite{Yeganefard2010EventBGuide}. Our case study has the potential to be elaborated into such guidelines. Although the effectiveness of refinement strategies has been discussed qualitatively in most cases, there have been efforts on quantitative analysis~\cite{Kobayashi2018RefinementAnalysis}. Our future work will include analysis of refinement strategies in this work in a more systematic way.

\section{Conclusion}
\label{sec:conclusion}
In this paper, we reported our case study to model, derive, and prove the safety conditions of AVs in the RSS scheme. We target a state-of-the-art problem with the goal-aware version of RSS as well as the Simplex architecture to consider black-box AI controllers. We proposed a strategy for leveraging the refinement mechanism of Event-B and demonstrated how it mitigates the complexity over scenario variations. We will continue studying other scenarios to convert the obtained lessons into more concrete and general guidelines for formal modelling and verification of AVs.

\section*{Acknowledgements}
We thank our industrial partner Mazda for discussions of realistic problems in the safety assurance of autonomous driving. We also thank members of JST ERATO HASUO Metamathematics for Systems Design Project for discussions of Goal-Aware RSS and the safety architecture. 

\bibliographystyle{splncs04}
\bibliography{main}

\end{document}